\documentclass[aps,prb,groupedaddress,twocolumn,showpacs]{revtex4}
\usepackage{epsfig}
\begin{document}

\title{Kondo lattice model
with a direct exchange interaction between localized moments}
\author{Mun Dae  Kim$^{1}$, Chul Koo Kim$^{2}$, and Jongbae Hong$^{1}$}
\affiliation{$^1$ School of Physics, Seoul National University, Seoul 151-742, Korea\\
$^{2}$ Institute of Physics and Applied Physics, Yonsei University, Seoul 120-749, Korea}

\begin{abstract}
We study the Kondo lattice model with a direct antiferromagnetic
exchange interaction between localized moments.
Ferromagnetically long-range ordered state coexisting with the Kondo screening shows
a continuous quantum phase transition to the Kondo singlet state.
We obtain the value of the critical point where the magnetizations of the localized moments and
the conduction electrons vanish.
The magnetization curves  yield a universal critical exponent independent
of the filling factors and the strength of the interaction between localized moments.
It is shown that the direct exchange interaction between localized moments introduces
another phase transition from an antiferromagnetic ordering
to a ferromagnetic ordering  for small Kondo exchange interaction.
We also explain the local minimum of the Kondo temperature in recent experiments.
\end{abstract}

\draft
\maketitle

\section{Introduction}

Interaction between itinerant electrons and  localized moments
in the Kondo regime is described by the s-f mixing and the polarization terms
in the Kondo lattice model.
The polarization term induces a RKKY type
magnetic correlation, while the s-f mixing causes demagnetization.
The magnetic correlation is dependent on the filling factor;
ferromagnetic for low filling and antiferromagnetic
around half filling. \cite{Lacroix}
As the interaction between the conduction electrons and the localized moments
increases, the phase evolves to the Kondo singlet phase.
In particular, for half-filled case,  the phase transition has been shown continuous
and the value of critical point has been obtained. \cite{Shi,Assaad,Jurecka,Zhang}

In reality, however, the direct exchange interactions between
localized moments should be included for a more comprehensive
understanding. The Kondo exchange interaction, $J_K$, between the
conduction electrons and the localized moments invokes the
effective exchange interaction between the localized moments, the
RKKY interaction $J_{\rm RKKY}\sim |J_K|^{2}$.  \cite{Iglesias,Rupp,Bernhard}
In the present study, in addition to the RKKY interaction,
we introduce a direct antiferromagnetic exchange interaction between localized moments.
In the ferromagnetic Kondo lattice model for manganites,
its effects have been widely investigated.  \cite{Yunoki}
In this study, we aim to address the effect of antiferromagnetic exchange
interaction, $J_{\rm af}$, on the phase diagram and on the critical
behavior of Kondo lattice model (KLM) for low filling cases
by calculating the density of states in a  mean field level.

First of all, we investigate the magnetizations of both
the lattice magnetic moments $\mu'$ and the conduction
electrons $\mu$ for various filling factors and various values of $J_{\rm af}$.
The magnetic moments $\mu'$and $\mu$ show qualitatively different behaviors;
$\mu'$ falls monotonously from its maximum value as $J_K$ increases,
whereas $\mu$ starts from zero, passes through  a local maximum,
and falls to zero at the same point as $\mu'$ does.
Near the critical point, the  magnetization curves show characteristic behaviors of
continuous phase transition. The critical behavior
of the magnetizations shows a universal exponent
independent of the filling factor and the strength of exchange interaction
$J_{\rm af}$ such that $\mu' (\mu) \sim (J_K -J_c)^{1/2}$ with the critical point $J_c$.
There emerge two mean fields. The first one, $x$,  represents the mean field of the s-f mixing
and the other field, $B$, the magnetic correlation between localized moments.
As $J_K$ increases,  the $x$ field increases whereas the $B$ field decreases,
which indicates a phase transition from the magnetically ordered state to the Kondo state.
Calculated values of the mean fields $x$ and $B$ explain consistently the the physical situation
such as the existence of phase transition between
the ferromagnetic ordered state and the Kondo singlet state.

A notable feature of the present calculation is the existence of a new antiferromagnetic
state at small Kondo exchange interaction especially for low filling factors.
The s-f mixing  field, $x$, is generally believed to be a monotonously
increasing function of $J_K$. \cite{Lacroix,Iglesias}
However, for very small $J_K$, we find a local minimum of $x$ field.
It is accompanied by  suppression of the $B$ field,
which is the precursor of a transition to another magnetic order,
an antiferromagnetic phase.
A recent  experiments on CeRu$_2$Ge$_2$ by S{\"u}llow {\it et al.}  \cite{Sullow}
and on CeAgSb$_2$ by Sidorov {\it et al.} \cite{Sidorov}
can be explained in connection with the anomalous behavior of the $x$ field and the $B$ field.

\section{Coexistence of magnetic ordering and Kondo screening}

The  magnetic correlation and the s-f mixing process are the two
competing orders in the KLM.  The Doniach diagram \cite{Doniach} shows that the former has
dominant effect for small $J_K$ while the latter for large $J_K$.
The Kondo exchange model includes the s-f mixing term and the polarization term
in the Kondo interaction to describe the competing orders.
The polarization term give rise to the RKKY interaction. \cite{Lacroix}

Integrating out the fast mode in the action including the Kondo interaction
may produce the effective RKKY interaction term in the Hamiltonian \cite{Neto}
with the interaction strength dependent on the Kondo interaction and the cutoff frequency.
In this study, however, we consider the case that the RKKY interaction
is  implicitly included in the polarization term of the Kondo Hamiltonian.
Instead we introduce the direct antiferromagnetic exchange
interaction between the localized moments whose strength we treat as independent parameter.
For ferromagnetic Kondo lattice model for manganites, where the
interaction between the conduction electrons and the localized
electrons is ferromagnetic, the effect of the
direct exchange interaction between localized moments has already been studied. \cite{Yunoki}
Here, various antiferromagnetic phases appear,
even though the antiferromagnetic exchange interaction
is  much weaker than that  of the Hund coupling. \cite{Ishihara,Horsch}

The Hamiltonian $H=H_0+V_K+V_H$ with
the Kondo interaction $V_K$ and the interaction between localized moments $V_H$
is written by
\begin{eqnarray}
H_0&=&\sum_{{\bf k}\sigma}\epsilon_{\bf k} c_{{\bf k}\sigma}^\dagger c_{{\bf k}\sigma}
+\sum_{{\bf i},\sigma}E_0 f_{{\bf i},\sigma}^\dagger f_{{\bf i},\sigma}\\
V_K&=&-J_K\sum_{\bf i} \left({\bf S}_{\bf i}\cdot {\bf s}_{\bf i}-\frac14 n_{f,{\bf i}} n_{c,{\bf i}}\right)\\
\label{VH}
V_H&=&-\frac12J_{\rm af}\sum_{{\bf i },{\bf a}}
\left({\bf S}_{\bf i}\cdot {\bf S}_{{\bf i}+{\bf a}}
-\frac14 n_{f,{\bf i}} n_{f,{\bf i}+{\bf a}}\right),
\end{eqnarray}
where $\sigma$ represents spin, {\bf a}  a unit vector connecting the nearest neighbors,
and $E_0$  the energy of the localized level.
$J_K (<0)$ is the  antiferromagnetic  Kondo exchange interaction and
$J_{\rm af} (<0)$ the antiferromagnetic exchange interaction between localized moments.
$c^\dagger_{k\sigma}$ is the creation operator for conduction electrons
with wave vector $k$, $f^\dagger_{i\sigma}$ for localized electrons at cite
$i$, $n_{c,i}=\sum_\sigma c^\dagger_{i\sigma}c_{i\sigma}$,
and $n_{f,i}=\sum_\sigma f^\dagger_{i\sigma}f_{i\sigma}$.

Using the Stratonovitch-Hubbard transformation and
the identities, ${\bf s}_{\bf i}=\frac12 c_{\bf i}^\dagger{\bf \sigma} c_{\bf i}$
and ${\bf S}_{\bf i}=\frac12 f_{\bf i}^\dagger{\bf \sigma} f_{\bf i}$,
with Pauli matrices ${\bf \sigma}$,   the interaction terms are represented by
$V_K=H_1+H_2-(J_K/2)\sum_{\bf i} x^2_{\bf i}$
and $V_H=H_3+H_4-(J_{\rm af}/4)\sum_{\bf i,a} y^2_{\bf i}$, where
\begin{eqnarray}
\label{H1}
H_1&=&\frac{J_K}2 \sum_{{\bf i},\sigma}x_{\bf i}(f^\dagger_{{\bf i},\sigma}c_{{\bf i},\sigma}
+c^\dagger_{{\bf i},-\sigma}f_{{\bf i},-\sigma})\\
\label{H2}
H_2&=&\frac{J_K}2\sum_{{\bf i},\sigma}[
\langle f^\dagger_{{\bf i},-\sigma}f_{{\bf i},-\sigma}\rangle c^\dagger_{{\bf i},\sigma}c_{{\bf i},\sigma}
+\langle c^\dagger_{{\bf i},\sigma}c_{{\bf i},\sigma}\rangle f^\dagger_{{\bf i},-\sigma}f_{{\bf i},-\sigma}
\nonumber\\
&-&\langle f^\dagger_{{\bf i},-\sigma}f_{{\bf i},-\sigma}\rangle \langle
c^\dagger_{{\bf i},\sigma}c_{{\bf i},\sigma}\rangle]\\
\label{H3}
H_3&=&\frac{J_{\rm af}}4 \sum_{{\bf i,a},\sigma}y_{\bf i}(f^\dagger_{{\bf i},\sigma}f_{{\bf i}+{\bf a},\sigma}
+f^\dagger_{{\bf i}+{\bf a},-\sigma}f_{{\bf i},-\sigma})\\
\label{H4}
H_4&=&\frac{J_{\rm af}}4\sum_{{\bf i},{\bf a},\sigma}[
\langle f^\dagger_{{\bf i},\sigma}f_{{\bf i},\sigma}\rangle
f^\dagger_{{\bf i}+{\bf a},-\sigma}f_{{\bf i}+{\bf a},-\sigma}\nonumber\\
&+&f^\dagger_{{\bf i},\sigma}f_{{\bf i},\sigma}
\langle f^\dagger_{{\bf i}+{\bf a},-\sigma}f_{{\bf i}+{\bf a},-\sigma}\rangle \nonumber\\
&-&\langle f^\dagger_{{\bf i},\sigma}f_{{\bf i},\sigma}\rangle
\langle f^\dagger_{{\bf i}+{\bf a},-\sigma}f_{{\bf i}+ {\bf a},-\sigma}\rangle],
\end{eqnarray}
where $x_{\bf i}$ and $y_{\bf i}$ are the mean fields representing
the s-f mixing and the magnetic correlation between localized moments
respectively. \cite {Iglesias,Rupp,Bernhard,Coqblin}
Hereafter, we consider the uniform case such that $x_{\bf i}=x$ and $y_{\bf i}=y$.

Here, $H_1$ describes the Kondo scattering and $H_2$ the polarization of the
conduction electrons by the localized moments. $H_2$ gives the magnetic
ordering through the RKKY interaction between localized moments.
The competition between $H_1$ and $H_2$ is previously studied \cite{Lacroix}.
$H_3$ comes from the transverse part of the direct exchange interaction $V_H$
in Eq. (\ref{VH}) and can be represented in k-space,  \cite{Iglesias}
\begin{eqnarray}
H_3=B\sum_{\bf k,\sigma}\epsilon_{\bf k} f^\dagger_{{\bf k}\sigma}f_{{\bf k}\sigma},
\end{eqnarray}
where $B\equiv -zyJ_{\rm af}/2D$ and $\epsilon_{\bf k}\equiv
-(D/z) \sum_{\bf a}\cos{\bf k}\cdot {\bf a}$ with coordination number $z$.
This term gives the localized moments  a small band width $2BD$. $H_4$ is the
term polarizing the nearest neighbor  localized moments.

In order to describe the critical behavior near the critical point,
we consider the coexistence of the magnetic ordering and the Kondo screening.
Calculating  magnetizations of the  localized moments and the conduction
electrons for various filling factors and various $J_{\rm af}$,
we  obtain the phase boundaries and the critical exponent.
In this case we have the relations such that
\begin{eqnarray}
\label{self1}
\langle c^\dagger_{{\bf i}\downarrow}c_{{\bf i}\downarrow}\rangle=\frac12(n+\mu),~~~
\langle c^\dagger_{{\bf i}\uparrow}c_{{\bf i}\uparrow}\rangle=\frac12(n-\mu),\\
\label{self2}
\langle f^\dagger_{{\bf i}\downarrow}f_{{\bf i}\downarrow}\rangle=\frac12(1-\mu'),~~~
\langle f^\dagger_{{\bf i}\uparrow}f_{{\bf i}\uparrow}\rangle=\frac12(1+\mu'),
\end{eqnarray}
where $\mu$ $(\mu')$ is the magnetization of s (f) electrons
in the spin-down (spin-up) direction and $n$ is the filing factor.
Since the Kondo exchange interaction
$J_K$ is antiferromagnetic, we expect that $\mu$ and $\mu'$ have the same sign.

With these equations the Green's function for the f-electrons and the c-electrons
corresponding to the Hamiltonian $H$ can be obtained as follows,
\begin{eqnarray}
\label{Green1}
&&G_{f\uparrow}(\omega)=\sum_k\left[\omega \pm is-E_0-B\epsilon_k-\frac14 J_K(n+\mu)-\right.\nonumber\\
&&\left.\frac14 zJ_{\rm af}(1-\mu')
-\frac{J^2_K x^2}{4[\omega \pm is-\epsilon_k-\frac14 J_K(1-\mu')]}\right]^{-1},\nonumber\\
&&\\
\label{Green2}
&&G_{c\uparrow}(\omega)=\sum_k\left[\omega \pm is-\epsilon_k -\frac14 J_K(1-\mu')-\right.\nonumber\\
&&\left.\frac{J_K^2 x^2}{4[\omega \pm is-E_0-B\epsilon_k
-\frac14 J_K(n+\mu)-\frac14 zJ_{\rm af}(1-\mu')]}\right]^{-1}.\nonumber\\
&&
\end{eqnarray}
For the spin down case, the quantities such as
$G_{f\downarrow}(\omega)$ can be simply obtained by substituting
$\mu\rightarrow -\mu$ and $\mu'\rightarrow -\mu'$.

The above Green's functions  can be arranged to be written as follows,
\begin{eqnarray}
G_{f\uparrow}(\omega)&=&\sum_k\left[\frac{P_{f\uparrow}(\omega)}{g_{f\uparrow}(\omega) \pm is -\epsilon_k}
+\frac{Q_{f\uparrow}(\omega)}{h_{f\uparrow}(\omega) \pm is-\epsilon_k}\right],\nonumber\\
\\
G_{c\uparrow}(\omega)&=&\sum_k\left[\frac{P_{c\uparrow}(\omega)}{g_{c\uparrow}(\omega) \pm is -\epsilon_k}+
\frac{Q_{c\uparrow}(\omega)}{h_{c\uparrow}(\omega) \pm is-\epsilon_k}\right].\nonumber\\
\end{eqnarray}
Here we use the flat band for the conduction electrons such that
${\rm Im}\sum_k 1/(\omega\pm is -\epsilon_k)=1/2D (0)$ for $-D<\omega<D$ (otherwise).
Hence, the density of states,
$\rho_{f\uparrow}(\omega)={\rm Im} G_{f\uparrow}(\omega)$,
can be obtained by
$\rho_{f\uparrow}(\omega)= P_{f\uparrow}(\omega)/2D$ $(Q_{f\uparrow}(\omega)/2D)$
for $\omega_{f1\uparrow}<\omega<\omega_{f2\uparrow}$
$(\omega_{f3\uparrow}<\omega<\omega_{f4\uparrow})$.
The density of states for the conduction electrons,
$\rho_{c\uparrow}(\omega)={\rm Im} G_{c\uparrow}(\omega)$,
can be obtained in a similar fashion.
Since we can readily show  $P_{f\uparrow}(\omega)=Q_{f\uparrow}(\omega)$
and $P_{c\uparrow}(\omega)=Q_{c\uparrow}(\omega)$ in the above, the density of states,
$\rho_{f\uparrow}(\omega)=\tilde{\rho}_{f\uparrow}(\Omega_\uparrow)$ and
$\rho_{c\uparrow}(\omega)=\tilde{\rho}_{c\uparrow}(\Omega_\uparrow)$, are given by
\begin{eqnarray}
\label{DOSf}
\tilde{\rho}_{f\uparrow}(\Omega_\uparrow)&=&
\frac1{4BD}\left[1+\frac{\Omega_{\uparrow}}{\sqrt{\Omega^2_{\uparrow}+BJ^2x^2}}\right]\\
\label{DOSc}
\tilde{\rho}_{c\uparrow}(\Omega_\uparrow)&=&
\frac1{4D}\left[1-\frac{\Omega_{\uparrow}}{\sqrt{\Omega^2_{\uparrow}+BJ^2x^2}}\right],
\end{eqnarray}
where
\begin{eqnarray}
\label{DefOmega}
\Omega_\uparrow (\omega) &= &(1-B)\omega-\frac14 J((n+\mu)-(1-\mu')B)\nonumber\\
&-&E_{0\uparrow}-\frac14zJ_{\rm af} (1-\mu').
\end{eqnarray}

The band edges can be obtained by solving
the equations $-D<g_{f (c)\uparrow}(\omega)<D$ and $-D<h_{f (c)\uparrow}(\omega)<D$.
Using the  expressions obtained for $g_\uparrow(\omega)$ and $h_\uparrow(\omega)$,
\begin{eqnarray}
g_\uparrow(\omega)&=&\frac1{2B} \left[(1+B)\omega-\frac14
J((n+\mu)+(1-\mu')B)\right.\nonumber\\
&-&E_{0\uparrow}-\frac14zJ_{\rm af}(1-\mu')
+\left.\sqrt{\Omega^2_{\uparrow}+J^2x^2B}\right]\\
h_\uparrow(\omega)&=&\frac1{2B} \left[(1+B)\omega-\frac14
J((n+\mu)+(1-\mu')B) \right.\nonumber\\
&-&E_{0\uparrow}-\frac14 zJ_{\rm af}(1-\mu')
-\left.\sqrt{\Omega^2_{\uparrow}+J^2x^2B}\right],
\end{eqnarray}
we get for $\omega_{i\uparrow}$
\begin{eqnarray}
\omega_{1\uparrow}&=&\frac12 [\omega_a-(1+B)D \nonumber\\
&-&\sqrt{\{\omega_b+(1-B)D\}^2-4BD^2+J^2x^2}]\\
\omega_{2\uparrow}&=&\frac12 [\omega_a+(1+B)D \nonumber\\
&-&\sqrt{\{\omega_b-(1-B)D\}^2-4BD^2+J^2x^2}]\\
\omega_{3\uparrow}&=&\frac12 [\omega_a-(1+B)D \nonumber\\
&+&\sqrt{\{\omega_b+(1-B)D\}^2-4BD^2+J^2x^2}]\\
\omega_{4\uparrow}&=&\frac12 [\omega_a+(1+B)D \nonumber\\
&+&\sqrt{\{\omega_b-(1-B)D\}^2-4BD^2+J^2x^2}],
\end{eqnarray}
where $\omega_a =\frac14 J((n+\mu)+(1-\mu'))+E_{0\uparrow}+\frac14 zJ_{\rm af}(1-\mu')$
and $\omega_b =\frac14 J((n+\mu)-(1-\mu'))+E_{0\uparrow}+\frac14 zJ_{\rm af}(1-\mu')$.
Here, we used the relation,
$g_{f\sigma}(\omega)=g_{c\sigma}(\omega)=g_\sigma(\omega)$,
$h_{f\sigma}(\omega)=h_{c\sigma}(\omega)=h_\sigma(\omega)$, and
$\omega_{fi\uparrow}=\omega_{ci\uparrow}=\omega_{i\uparrow}$.

\section{Results and discussions}

The ferromagnetic magnetizations $\mu'$ and $\mu$ in presence of the Kondo screening
can be obtained as a function of the Kondo exchange interaction $J_K$
by solving the self consistent equations,  Eqs. (\ref{self1}) and (\ref{self2}),
using the density of state,  Eqs. (\ref{DOSf}) and (\ref{DOSc}).
Eqs. (\ref{self1}) and (\ref{self2}) for spin up case are represented   as
\begin{eqnarray}
\label{self3}
\frac{1+\mu'}2=\frac1{1-B}\int^{\Omega_{F\uparrow}}_{\Omega_{1\uparrow}}\tilde{\rho}_{f\uparrow}(\Omega) d\Omega\\
\label{self4}
\frac{n-\mu}2=\frac1{1-B}\int^{\Omega_{F\uparrow}}_{\Omega_{1\uparrow}}\tilde{\rho}_{c\uparrow}(\Omega) d\Omega.
\end{eqnarray}
From the  above equations $\Omega_{F\uparrow}$ and $\Omega_{1\uparrow}$ can be evaluated
by inserting the density of states from Eqs. (\ref{DOSf}) and (\ref{DOSc}).
$\Omega_\uparrow(E_F)$ and $\Omega_\uparrow(\omega_{1\uparrow})$ can also be
represented from Eq. (\ref{DefOmega}). Solving the equations,
$\Omega_{F\uparrow}=\Omega_\uparrow(E_F)$ and
$\Omega_{1\uparrow}=\Omega_\uparrow(\omega_{1\uparrow})$,
$E_{0\uparrow}$ and $E_{F\uparrow}$ can be given by
\begin{eqnarray}
\label{E0}
&&E_{0\uparrow}=\nonumber\\
&&\frac{4\Omega^2_{1\uparrow}+(1-B)^2(4BD^2-J^2x^2)}
{-2(1+B)\Omega_{1\uparrow}+(1-B)\sqrt{4\Omega^2_{1\uparrow}-16B^2D^2+4BJ^2x^2}}\nonumber\\
&&-\frac14 J((n+\mu)-(1-\mu'))-\frac14 zJ_{\rm af}(1-\mu')-(1-B)D\nonumber\\
\\
\label{Ef}
&&E_{F\uparrow}=\frac1{1-B} \left[\Omega_{F\uparrow}+E_{0\uparrow}\right.\nonumber\\
&&+\left.\frac14 J((n+\mu)-(1-\mu')B)+\frac14 zJ_{\rm af}(1-\mu')\right].
\end{eqnarray}

The values of magnetizations, $\mu'$ and $\mu$, are obtained numerically.
First, we fix the values of $x$ and $B$ for given values of $n$, $J_{\rm af}$, and $J_K$
and, then, solve the equations,
$E_{F\uparrow}=E_{F\downarrow}$ and $E_{0\uparrow}=E_{0\downarrow}$,
to obtain the values of magnetizations, $\mu'$ and $\mu$.
From these values we calculate the total energy of the conduction electrons and localized moments.
The total energy can be written as
$E=\sum_\sigma \int_{\omega_{1\sigma}}^{E_F} \omega
[\rho_{f\sigma}(\omega)+\rho_{c\sigma}(\omega)]d\omega+E_{MF}$.
$E_{MF}$ can be obtained from the Hamiltonians in Eqs. (\ref{H1})-(\ref{H4}) with the
mean values from Eqs. (\ref{self1}) and (\ref{self2}) and is given by
$E_{MF}=-(1/2)J_K x^2-BD^2/zJ_{\rm af} -(1/4)J_K (n+\mu\mu')-(1/8)zJ_{\rm af}(1-\mu'^2)$.
For appropriate ranges of  $x$ and $B$ values, we repeat the above process and
determine the values of $\mu'$ and $\mu$ at the minimum point of the total energy of the system.

\begin{figure}[t]
\vspace{4cm}
\includegraphics{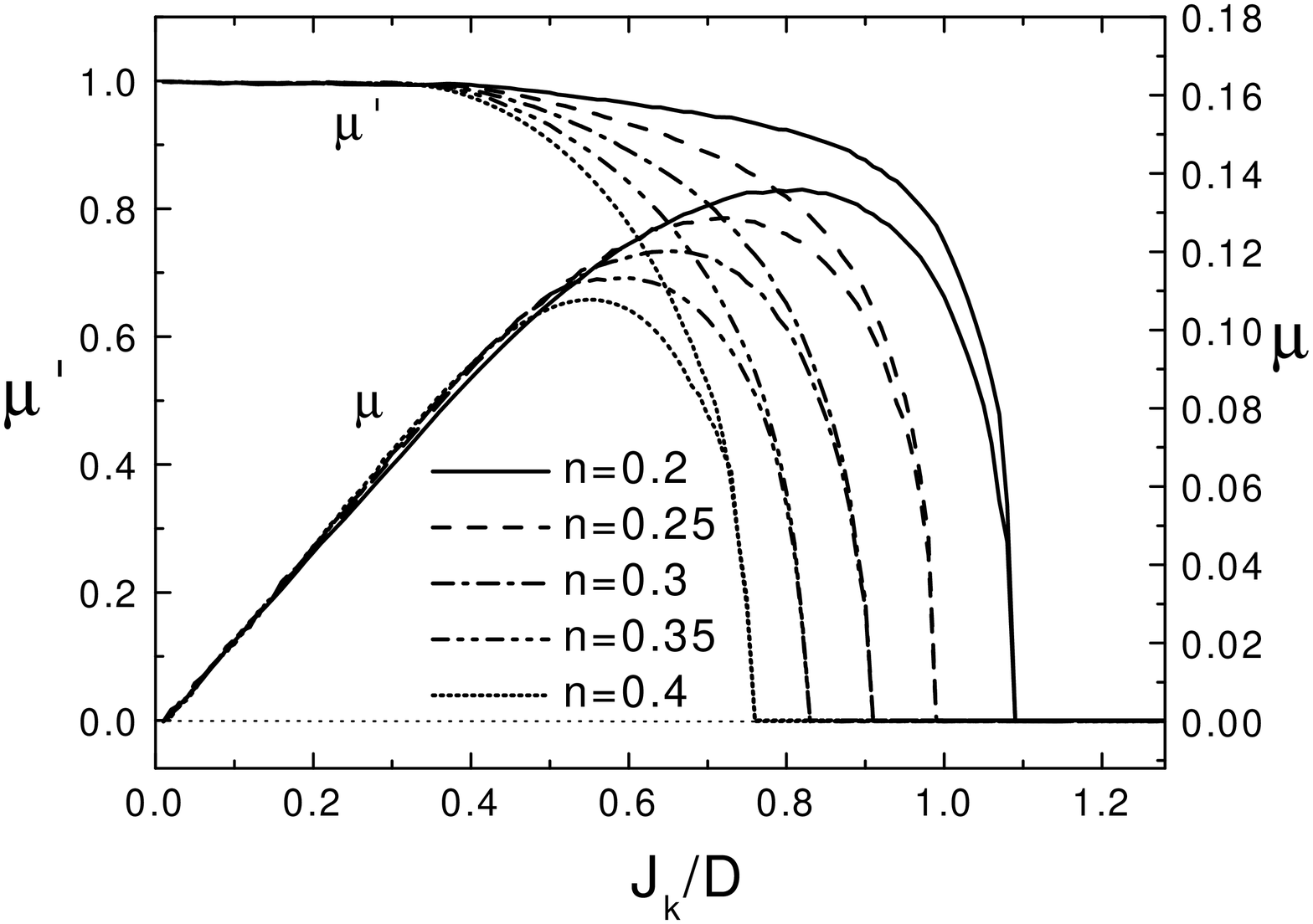} \vspace{2cm} \caption{Magnetization of the localized
moments, $\mu'$ and of the conduction electrons, $\mu$ in
ferromagnetic state for various filling factors, $n$, as functions
of the Kondo exchange interaction strength, $J_K$. Here, we set
the strength of direct exchange interaction $J_{\rm
af}/D=-0.001$.} \label{fig:Mm}
\end{figure}

\begin{figure}[b]
\vspace{6cm} \includegraphics{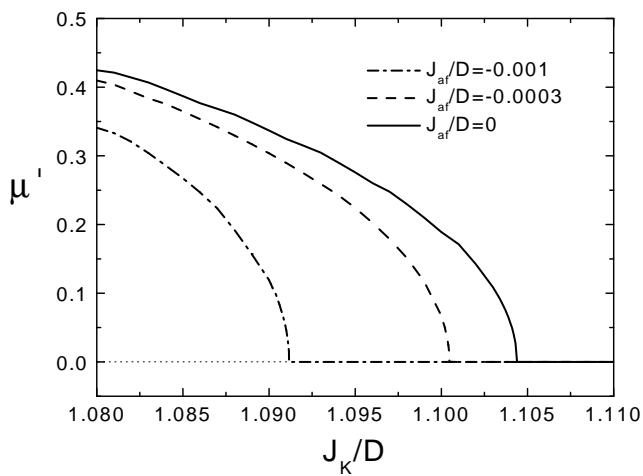} \vspace{0.5cm}
\caption{Magnetization of the localized moments, $\mu'$ for
various values of direct exchange interaction $J_{\rm af}$ between
localized moments with a filling factor, $n=0.2$.}
\label{fig:Mn20}
\end{figure}

In Fig. \ref{fig:Mm}, we show the  magnetization of the localized moments, $\mu'$, and
of the conduction electrons, $\mu$, in ferromagnetic state
when $J_{\rm af}/D=-0.001$. The magnetization curves  clearly demonstrate existence of  continuous quantum
phase transition. We observed that the critical values of the Kondo exchange interaction are higher
for  smaller values of filling factors, which is consistent
with previous studies. \cite{Lacroix}
However, there have been conflicting reports on  the behavior of $\mu$.
A recent Monte Carlo study \cite{Assaad} showed existence of  a local maximum in the $\mu$
curve, but its behavior for small values of $J_K$ was left uncertain.
On the contrary, a more recent bond-operator mean-field study \cite{Jurecka}
produced a monotonously decreasing behavior. Present study shows that
the magnetization of conduction electrons $\mu$ starts to increase from zero and pass a local maximum
before it falls to zero at the same points as those of $\mu'$. It is consistent with the Monte
Carlo calculation and also clarifies the situation for small values of $J_K$.

\begin{figure}[t]
\vspace{7cm} \includegraphics{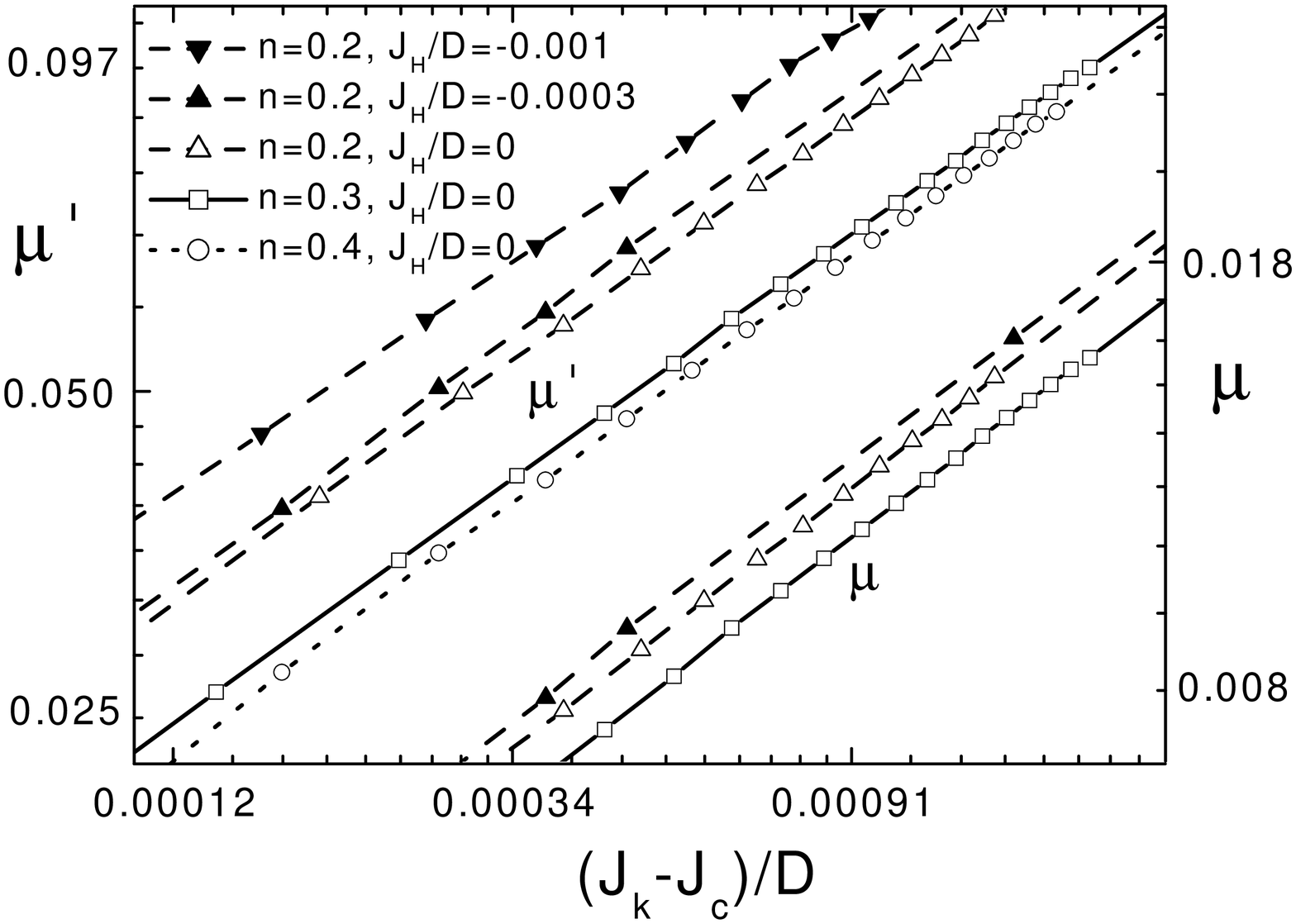} \vspace{-0.5cm} \caption{Log-log
plot of the magnetization of localized moments $\mu'$ and
conduction electrons $\mu$ near the critical point $J_c$.}
\label{fig:MmScale}
\end{figure}

Fig. \ref{fig:Mn20} shows the behavior of $\mu'$ near the critical
point for $n$=0.2 with various values of $J_{\rm af}$. $J_{\rm
af}$ competes with the ferromagnetic correlation and, thus, makes
the critical points slightly smaller as shown in Fig.
\ref{fig:Mn20}. We observe that the critical points are dependent
on the values of $J_{\rm af}$ as well as the filling factors.
Close study on the critical behaviors of  $\mu'$ and $\mu$
clearly shows that the critical exponent has a universal value
independent of the filling factor and $J_{\rm af}$.
The form is given by $(J_K-J_c)^{1/2}$ as shown in Fig. \ref{fig:MmScale}.

\begin{figure}[b]
\vspace{12cm} \includegraphics{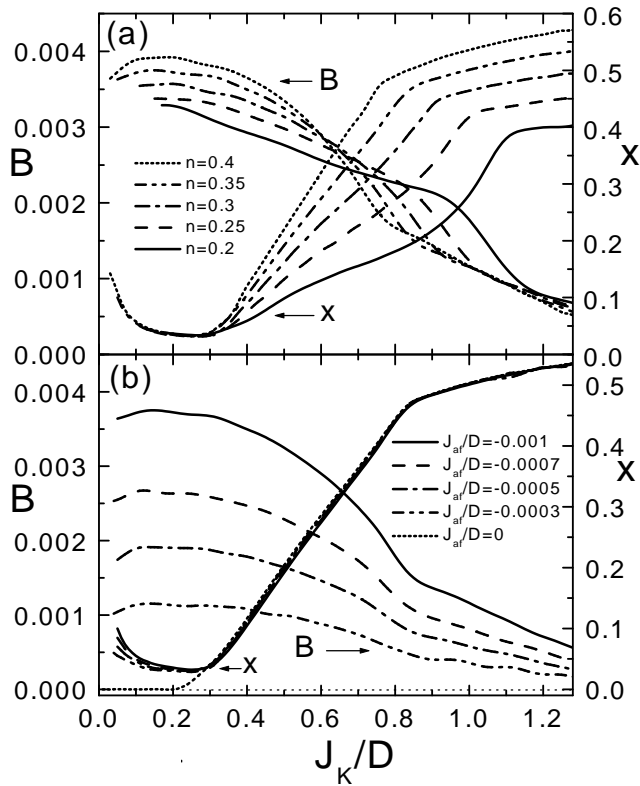} \vspace{-1cm}
\caption{ Plots for the mean field $x$ and $B$ (a) for various $n$ values when
$J_{\rm af}/D=-0.001$ and  (b) for various values of $J_{\rm af}/D$ when $n=0.35$.}
\label{fig:Bx}
\end{figure}

The mean field variable $x$ representing the s-f mixing
is closely related to the Kondo temperature
and the variable $B$ representing the  correlation between
localized spins is to the correlation temperature.  \cite{Iglesias}
The behaviors of $x$ and $B$ as  functions of $J_K$
are plotted in Fig. \ref{fig:Bx}(a).
Fig. \ref{fig:Bx}(a) shows that both $x$ and $B$ have  higher values
for higher filling factors, which is consistent with the study by Ruppenthal et al. \cite{Rupp}
The field $B$ is shown to behave  in the opposite direction of $x$.
This signifies that
$B$ represents the magnetically ordered state competing with the Kondo singlet state.

\begin{figure}[t]
\vspace{7cm} \includegraphics{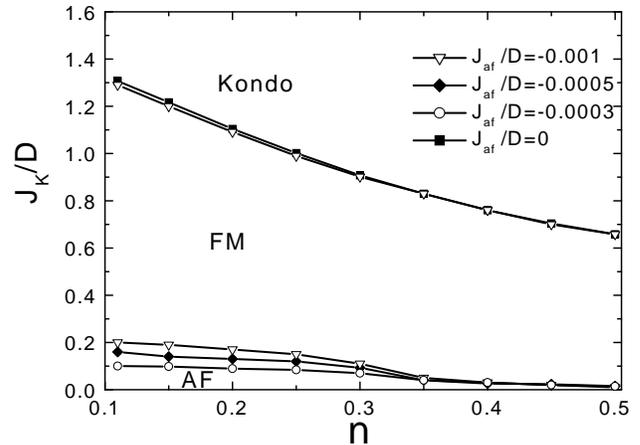} \vspace{-0.5cm} \caption{Phase
diagram for various $J_{\rm af}$. The phase boundary for the
antiferromagnetic state and the ferromagnetic state is shown in
the lower part of the figure. The phase boundary for the
ferromagnetic state and the Kondo singlet state is shown in the
upper part for $J_{\rm af}/D=0$ and $J_{\rm af}/D=-0.001$.}
\label{fig:diagram}
\end{figure}

The KLM with a direct exchange interaction between localized moments exhibits
a qualitatively different behaviors  from those without direct interaction
between localized moments. The value of $x$ shows a local minimum
as $J_K$ becomes quite small. This phenomenon is accompanied by the suppression  of the $B$ field.
This behavior indicates that the magnetic correlation
becomes weak   and another transition is being anticipated.
The local minimum of $x$ is shown in Fig. \ref{fig:Bx} (a)
for various filling factors when $J_{\rm af}/D=-0.001$ and
for various values of $J_{\rm af}$ when $n$=0.35 in (b), where
the re-growth of $x$ is shown stronger for larger $|J_{\rm af}|$.

We have  calculated  the  antiferromagnetic state energy and
compared it with that of the ferromagnetic state.
In Fig. \ref{fig:diagram}, the phase boundary for the ferromagnetic
state and the antiferromagnetic state is shown in the lower part of the figure.
For low filling, the antiferromagnetic state become more favorable.
As the value of $J_K$ increases, the transition to the Kondo
singlet state  follows  as shown in Fig. \ref{fig:diagram}.
Similar phase boundary has been obtained previously by
comparing the energy of the Kondo singlet state
with that of the ferromagnetic state without Kondo scattering. \cite{Lacroix}
Here, it should be noted that
the present calculation is carried out when the ferromagnetism and the Kondo
screening coexist.

This behavior explains the recent experiments on CeRu$_2$Ge$_2$ by S{\"u}llow et al.  \cite{Sullow}
and on CeAgSb$_2$ by Sidorov {\it et al.} \cite{Sidorov}
In these experiments, the interaction between the localized moments, Ce,
is ferromagnetic while the RKKY interaction is antiferromagnetic.
Two consecutive phase transitions have been observed:
a transition from the ferromagnetic phase
to the antiferromagnetic phase followed by a transition
to the Kondo singlet state as $J_K$ increases further.
If we  exchange the role of the antiferromagnetic interaction
with that of the  ferromagnetic interaction, we can explain the experiment in the
present scenario, since here we have considered the antiferromagnetic interaction between f-electrons and
the ferromagnetic RKKY interaction.

In the experiment on CeRu$_2$Ge$_2$ \cite{Sullow} we observe
suppression of the magnetic transition temperature
near  the critical point where the transition from one magnetic state to
another magnetic state occurs. This corresponds to
the suppression of the $B$ field in the present study.
We can see in the experiments the local minimum of the Kondo temperature, $T_K$.
In the experiment on CeRu$_2$Ge$_2$ \cite{Sullow} we can only observe that
the Kondo temperature resists going to zero as $J_K$ decreases.
However, in the experiments on CeAgSb$_2$ \cite{Sidorov} and on CeRhIn$_5$ \cite{Hegger},
we clearly see the local minimum of the Kondo temperature.
This local minimum has been conjectured to be caused by  a certain mechanism competing with
the Kondo effect. In the present theory, we show that the direct interaction between the
localized moments competing with the Kondo scattering gives rise to the local minimum  of the $x$  field
and, thus, the Kondo temperature.


\section{Conclusion}

We have studied the KLM with an antiferromagnetic exchange interaction
between localized moments for a general situation in which
ferromagnetic correlation coexists with the Kondo screening.
Calculating the magnetization of the localized moments and the
conduction electrons as a function of the Kondo interaction
for various filling factors and values of $J_{\rm af}$,
we obtained a continuous quantum phase transition
from the ferromagnetic ordered state to the Kondo singlet state and the
critical points where the magnetizations vanishes. Around the critical
points, we obtained  the critical exponent independent of the filling factor and the
values of $J_{\rm af}$ such that $\mu'$ ($\mu) \sim (J_K-J_c)^{1/2}$.

Furthermore, we obtained the behavior of the  mean fields, $x$ and $B$,
corresponding to the Kondo scattering  and the
magnetic correlation respectively. As the Kondo interaction increases
the value of $B$ decreases while $x$ increases,
which explains existence of the Kondo singlet
state at large value of $J_K$. On the contrary, as $J_K$ decreases, the value of
$B$ increases while $x$ decreases. Hence, the enhanced magnetic correlation
with suppressed Kondo screening leads the magnetically ordered state.

When $J_K$ is further reduced, the another phase transition from the antiferromagnetically
ordered state to the ferromagnetically appears due to the $J_{\rm af}$ interaction.
Comparing the energy of the ferromagnetic state and the antiferromagnetic state,
we obtain the phase boundary, which shows clearly that an antiferromagnetic state
appears for low filling.
The $x$ field exhibits a re-growth accompanied by the suppression of $B$.
This phenomena is due to the $J_{\rm af}$ interaction and
is a precursor for the transition to the antiferromagnetically
ordered state. Recent experiments \cite{Sullow,Sidorov,Hegger} can be explained
through the present theory.

\begin{center}
{\bf ACKNOWLEDGEMENTS}
\end{center}

This work was partly supported by Korea Research Foundation (2001-005-D20004)
and the BK21 project of the Ministry of Education, Korea.


\begin{references}
\bibitem{Lacroix} C. Lacroix and M. Cyrot, Phys. Rev. B. {\bf 20}, 1969 (1979).
\bibitem{Shi} Z.-P. Shi, R. R. P. Singh, M. P. Gelfand, and Z. Wang,
Phys. Rev. B {\rm 51}, 15630 (1995).
\bibitem{Assaad} F. F. Assaad, Phys. Rev. Lett. {\bf 83}, 796 (1999).
\bibitem{Jurecka} C. Jurecka and W. Brenig, Phys. Rev. B {\bf 64}, 092406 (2001).
\bibitem{Zhang} G.-M. Zhang and L. Yu, Phys. Rev. B {\bf 62}, 76 (2000).
\bibitem{Iglesias} J. R. Iglesias, C. Lacroix, and B. Coqblin,
Phys. Rev. B {\bf 56}, 11820 (1997).
\bibitem{Rupp} A. R. Ruppenthal, J. R. Iglesias, and M. A. Gusm{\~ a}o,
Phys. Rev. B {\bf 60}, 7321 (1999).
\bibitem{Bernhard} B. H. Bernhard, C. Lacroix, J. R. Iglesias, and B. Coqblin,
Phys. Rev. B {\bf 61}, 441 (2000).
\bibitem{Yunoki} S. Yunoki and A. Moreo, Phys. Rev. B {\bf 58}, 6403 (1998);
A. L. Malvezzi, S. Yunoki, and E. Dagotto, {\it ibid.} {\bf 59}, 7033 (1999).
\bibitem{Sullow} S. S{\" u}llow, M. C. Aronson, B. D. Rainford, and P. Haen,
Phys. Rev. Lett. {\bf 82}, 2963 (1999).
\bibitem{Sidorov} V. A. Sidorov, E. D. Bauer, N. A. Frederick, J. R. Jeffries,
S. Nakatsuji, N. O. Moreno, J. D. Thompson, M. B. Maple, and Z. Fisk,
Phys. Rev. B {\bf 67}, 224419 (2003).
\bibitem{Doniach} S. Doniach, Physica B {\bf 91}, 231 (1977).
\bibitem{Neto} A. H. Castro Neto and B. A. Jones, Phys. Rev. B {\bf 62}, 14975 (2000).
\bibitem{Ishihara}  S. Ishihara, J. Inoue, and S. Maekawa, Phys. Rev. B {\bf 55}, 8280 (1997).
\bibitem{Horsch} P. Horsch, J. Jakli$\check{c}$, and F. Mack, Phys. Rev. B {\bf 59}, R14149 (1999).
\bibitem{Coqblin} B. Coqblin, C. Lacroix, M. A. Gusm{\~a}o, and J. R. Iglesias,
Phys. Rev. B {\bf 67}, 064417 (2003).
\bibitem{Hegger} H. Hegger, C. Petrovic, E. G. Moshopoulou, M. F. Hundley, J. L. Sarrao,
Z. Fisk, and J. D. Thompson, Phys. Rev. Lett. {\bf 84}, 4986 (2000).
\end{references}
\end{document}